\documentclass{elsart}

\input{epsf}

\begin{document}

\begin{frontmatter}
 
\title{
       Evidence of microscopic effects in fragment mass distribution in 
       heavy ion induced fusion-fission reactions 
      }

\author[saha]{T.~K.~Ghosh},
\ead{tilak.ghosh@saha.ac.in, Tel : 91 33 23370379, Fax : 91 33 23374637}
\author[saha]{S.~Pal},
\author[nsc]{K.~S.~Golda} and
\author[saha]{P.~Bhattacharya}

\address[saha]{Saha Institute of Nuclear Physics,  1/AF  Bidhan  Nagar,
  Kolkata  700 064, India}
\address[nsc]{Nuclear Science Centre, New Delhi-110067, India}

\begin{abstract}

Our measurements of variances ($\sigma_{m}^2$) in mass distributions of 
fission fragments from fusion-fission reactions of light projectiles 
(C, O and F) on deformed thorium targets exhibit a sharp anomalous increase 
with energy near the Coulomb barrier, in contrast to the smooth variation 
of  $\sigma_{m}^2$ for the spherical bismuth target. This departure from 
expectation based on a statistical description is explained in terms of 
microscopic effects arising from the orientational dependence in the case 
of deformed thorium targets. 



\end{abstract}

\end{frontmatter}

\newpage

The formation of a super heavy element through primary fusion of two nuclei is 
restricted by the subsequent evolution of the compound system dependent on its 
survival (with at most particle emission) as opposed to its fission. The 
nuclei must have enough kinetic energy to overcome the repulsive electrostatic 
energy in order to come within the range of the attractive nuclear forces in 
a touching configuration. The path the system takes in a complicated 
multidimensional potential energy landscape \cite{NatureMollerNV03} governs 
the fusion of the two nuclei from a touching configuration to a composite 
system, equilibrated in all macroscopic degrees of freedom. As an example, 
depending upon the initial conditions of excitation, the entrant 
system of target and projectile can reach a fusion meadow in the energy 
landscape, equilibrate to a compound nucleus and cool down after the 
evaporation of a few particle and photon emission to a evaporation residue 
(ER), or the super-heavy compound nucleus could choose another path to undergo 
shape oscillations over an unconditional saddle to reach a fission valley.

The topography of the potential energy surface in the parameter space 
(involving the deformations of the two touching nuclei, their mass asymmetry, 
the separation between the two and the nature of the neck joining them) is far 
too complicated to enable us to determine theoretically the path taken by the 
system in its evolution. This is even more so, because of possible microscopic 
effects. Accordingly, it is of paramount importance to use experimental probes
 together with a phenomenological understanding to elucidate the route actually traversed by the system. Thus the observed angular anisotropy in the fission 
fragments (ratio of yields parallel and perpendicular to the beam direction), 
following statistical laws,  on one hand and the measured cross sections for 
production of evaporation residues on the other, are generally taken to 
indicate that the system equilibrates to compound nucleus in the fusion meadow. However, recent interpretations \cite{NishioPRL04,HindePRL95} based on 
measurement by these two probes have led to contradictory conclusion regarding  the path taken by the system vis a vis the fusion meadow or the fission valley 
or for that matter through an entirely different route over an asymmetric 
saddle. The present authors \cite{myRAPID2} have proposed that accurate 
measurements of mass distribution can be used as reliable tool to help pin 
down the route followed by the system to reach the fission valley. The 
present letter reports accurate measurements of fission fragment mass 
distributions as a function of the excitation energy close to the Coulomb 
barrier in several systems with different projectiles on a 
deformed as well as a spherical target and the phenomenological explanations of the observed variations of the width of mass distributions for different 
topographical routes the systems follow through the energy landscape. Our 
measurements for the first time clearly picks up the microscopic 
effects in determining the path the systems follow in reaching the fusion 
meadow or the fission valley.

The experiments were performed with judiciously chosen projectiles of  
$^{12}$C, $^{16}$O and $^{19}$F on deformed  $^{232}$Th and spherical $^{209}$Bi targets. Large deviations in the fragment anisotropy from the predictions of 
statistical theory \cite{HalpernStrutinsky} were reported for thorium target 
\cite{NMPRL96,RamPRL90,ZhangPRC94}, while those for spherical bismuth 
target followed the statistical 
predictions \cite{SamantEPJ00,KailasPhysRep}. For the spherical bismuth target, 
the entrant system is compact for any orientation and the expected mass 
flows are from target to projectile in all target-projectile 
systems \cite{Abe}. However, the compactness in shape for the entrance channel 
changes quite appreciably as the impact point of the projectile changes from 
the equatorial to the polar regions of the prolate thorium nuclei, and the 
macroscopic effects of mass flow for carbon ( projectile to target) is opposite to that of oxygen and fluorine nuclei (target to projectile) reacting with 
thorium target. In all the cases, the macroscopic effects only predict a 
smooth variation of the width of the fragment mass distributions with the 
excitation energies or the temperature of the equilibrated fused system 
\cite{PRC87Shen}. So any departure of the smooth variation of the width of 
the mass distributions would be a likely signature of microscopic effects 
driving the systems through different pathways in the energy landscape.

\begin{figure}[htb]
\centerline {\epsfxsize=3.0in,\epsffile{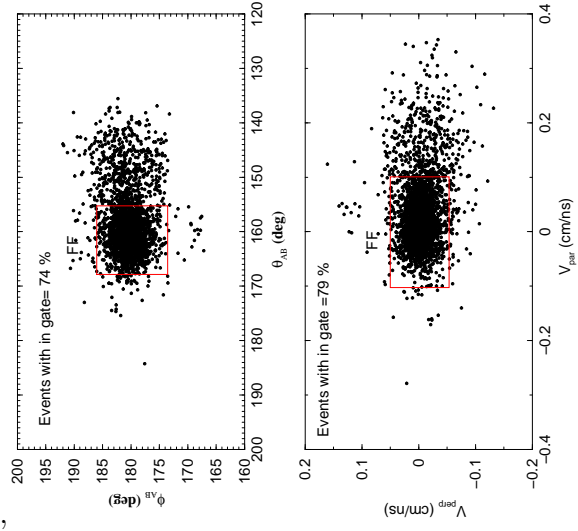} }
\vspace{0.15in}
\caption{Distributions of complementary fission fragments in 
$\theta$-$\phi$ (upper panel) and $V_{par}$-$V_{perp}$ (lower panel). The 
contour represents the gate used to select the fusion fission events. }
\label{fg:compare1}
\end{figure}

Pulsed heavy ion beams from the 15UD Pelletron at Nuclear Science Centre (NSC), New Delhi, India, had been used in the experiments. The pulse width was about 
0.8-1.5 ns with a pulse separation of 250 ns. The energy of the beams were 
varied typically in steps of 1-2 MeV, from a few MeV above the Coulomb barrier 
to a few MeV below it. The targets were either self-supporting $^{232}$Th of 
thickness 1.8 mg/cm$^2$ or a 500 $\mu$g/cm$^2$ thick self-supported 
$^{209}$Bi. Complementary fission fragments were detected with two large 
area (24 cm $\times$ 10 cm) X-Y position sensitive multi-wire proportional 
counters (MWPCs) \cite{myNIM04}. The fission fragments were separated from 
elastic and quasi-elastic channels using time of flight of particles and the 
energy loss signal in the detectors. Folding angle technique was 
used to differentiate between fusion-fission (FF) and transfer fission (TF) 
channels, from a distribution of the events in $\theta-\phi$ correlations 
or an equivalent procedure  of the correlation of the 
fissioning system velocities parallel and perpendicular ($V_{par}- V_{perp}$) 
to the reaction plane \cite{PB95}. In Fig.~\ref{fg:compare1}, typical 
separation of fragments from exclusively FF reactions are shown for both 
procedures. The resulting fragment mass widths differ at most few 
percent and clearly do not have any impact on the final experimental 
results or conclusions drawn from it. The masses of the fission 
fragments were determined event by event 
from precise measurements of flight paths and flight time differences of 
complementary fission fragments. The estimated mass resolution for 
fission fragment was about 3 a.m.u. The details of experimental arrangement 
and data analysis and elimination of systematic errors were reported in 
reference \cite{myNIM04,myRAPID1}.

\begin{figure}[htb]
\centerline {\epsfxsize=2.0in,\epsffile{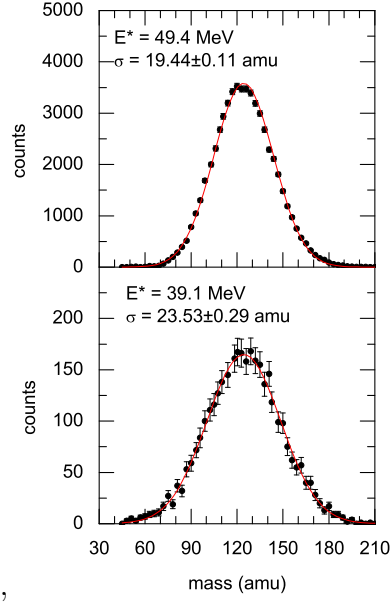} }
\vspace{0.15in}
\caption{Mass distributions at two excitation energies for the system 
$^{19}$F + $^{232}$Th. The Gaussian fit are shown by the solid lines.}
\label{fg:massdis}
\end{figure}

\begin{figure}[htb]
\centerline {\epsfxsize=4.5in \epsffile{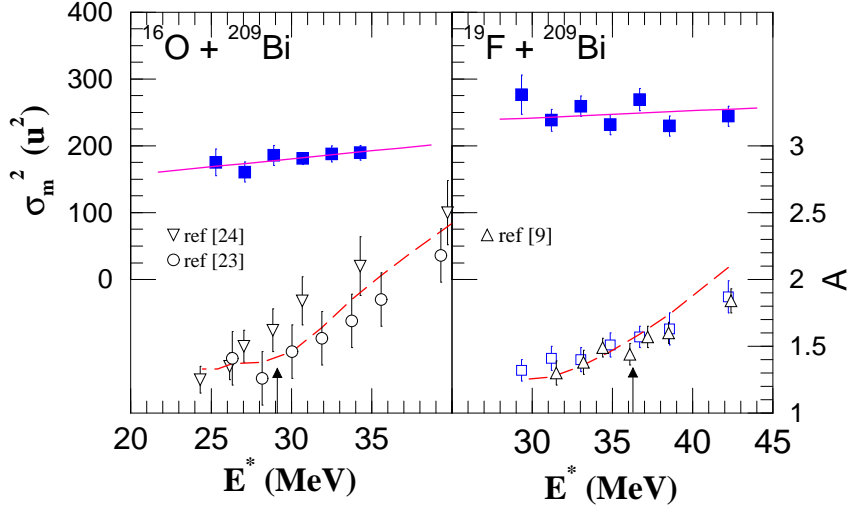} }
\vspace{0.15in}
\caption{Mass variance ($\sigma_{m}^2$) as a function of 
excitation energy (E$^\star$) for spherical bismuth target. The arrow points 
to excitation energy corresponding to Coulomb barrier. The solid lines 
 show smooth variation of $\sigma_{m}^2$ with E$^\star$. Reported fragment 
anisotropy A (open symbols) and SSPM predictions (dashed lines) are shown in 
lower halves.}
\label{fg:spherical}
\end{figure}

The measured mass distributions in earlier reported cases of $^{19}$F, 
$^{16}$O + $^{232}$Th and $^{16}$O +$^{209}$Bi \cite{myRAPID2,myRAPID1} and 
the presently measured case of $^{12}$C + $^{232}$Th and $^{19}$F + $^{209}$Bi 
at all energies are well fitted with single Gaussian distributions around the 
symmetric mass split for the target plus projectile systems. 
Typical mass distributions for the system $^{19}$F + $^{232}$Th, at excitation 
energies of 49.4 MeV and 39.1 MeV, fitted with a Gaussian  are shown in 
Fig.~\ref{fg:massdis}. The variation of  the variance of the fission fragment 
mass distribution ($\sigma_{m}^2$) are shown by solid squares in 
Fig.~\ref{fg:spherical} for $^{19}$F and $^{16}$O projectiles on the spherical 
$^{209}$Bi nuclei. It has been observed that the mass variance ($\sigma_{m}^2$) 
shows a smooth variation (trend is shown by solid lines) with the excitation 
energy of the fused system across the Coulomb barrier. This is in qualitative 
agreement with the predictions of statistical theories. It is also noted 
that no significant departures are reported in the fragment angular anisotropy
 measurements as shown by the open symbols in the lower halves of the 
figures (predicted anisotropies from SSPM theory \cite{HalpernStrutinsky} 
shown by dashed lines) for the spherical target and projectile systems 
\cite{SamantEPJ00,myRAPID2,KailasPhysRep}.Thus for 
these target-projectile combinations, we conclude that the systems fused to  
an equilibrated compound nucleus in the fusion meadow for all excitation 
energies, and subsequently underwent shape changes to reach an unconditional 
mass symmetric saddle and fission. Predominantly macroscopic forces are assumed 
to govern the paths taken by the above systems.

\begin{figure}[htb]
\centerline {\epsfxsize=5.0in \epsffile{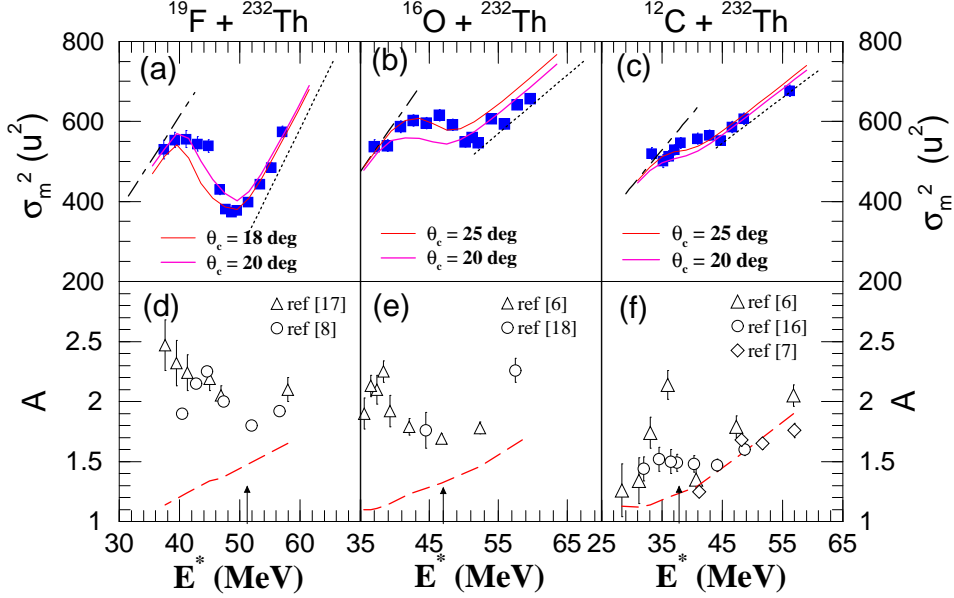} }
\vspace{0.15in}
\caption{Variation of $\sigma_{m}^2$ (solid squares) with 
excitation energy  for three systems. The dotted and dot-dashed curves are 
 variation for normal and postulated quasi-fission modes, respectively. 
Calculated $\sigma_{m}^2$ (thin and thick solid lines) are shown for two 
critical angles ($\theta_c$).Reported anisotropy A (open symbols) and 
SSPM predictions (dashed lines) are shown in lower panels .The arrow points 
to excitation energy corresponding to Coulomb barrier.}
\label{fg:deformed}
\end{figure}

The variances of mass distribution ($\sigma_{m}^2$) for reactions of different 
projectiles for the present as well as our earlier reports 
\cite{myRAPID1,myRAPID2} on the deformed thorium target are shown in 
upper panel of Fig.~\ref{fg:deformed} a-c, for $^{19}$F ,$^{16}$O, $^{12}$C 
projectiles, respectively. In all three cases, as the excitation energy is 
decreased, the $\sigma_{m}^2$ values, shown by solid squares, decreased 
monotonically, but shows a sudden upward trend approximately at the Coulomb 
barrier energies. This is once again followed by a smooth decrease as energy 
is further lowered. The sudden increase in $\sigma_{m}^2$ values is most 
prominent ($\sim 50\%$) in case of $^{19}$F + $^{232}$Th and decreases to 
$\sim ~ 15\%$ in $^{16}$O + $^{232}$ Th and to $\sim ~10\%$  in the 
$^{12}$C + $^{232}$ Th system. It has been simulated and experimentally 
verified that sudden rise in $\sigma_{m}^2$ values could not be explained  by 
any systematic error, e.g., loss of energy of fragments in target or mismatch 
of timing in two T.O.F. arms. The anomalous increase in angular anisotropy 
in all these systems  
\cite{NMPRL96,RamPRL90,ZhangPRC94,LestonePRC97,NMPRC95,BackJH} has been 
shown by open symbols in the lower panel of  Fig.~\ref{fg:deformed} d-f. 
It is interesting to note that anomalous increase in width of the 
mass distribution were observed at almost the same beam energies at 
which anomalous enhancement in fragment angular anisotropy were 
reported.

Observation of a sudden rise in $\sigma_{m}^2$ values as the excitation 
energy is lowered may signify a mixture of two fission modes, one following 
the normal statistical prediction of fusion-fission path along zero left-right mass asymmetry ($\alpha$), and another following a different path in the 
energy landscape with zero or small mass asymmetry. The mixture of the two 
modes could give rise to wider mass distributions. Similar to the postulation 
of the orientation dependent quasi-fission \cite{HindePRL95}, we postulate 
that for fusion-fission paths corresponding to the projectile orientations up 
to a critical angle ($\theta_c$) of impact on the polar region of prolate 
thorium, the width and energy slope of the symmetric mass distributions are 
different, as shown by dot-dashed curves in Fig.~\ref{fg:deformed} a-c, compared to those for the normal statistical fusion-fission paths (dotted curves). 
The mass widths weighted by the fission cross sections (which are assumed 
to be very close to fusion cross section as the composite systems are of 
high fissility) from earlier measurements \cite{NMPRL96} are mixed for the 
two fusion-fission modes and shown  by thick and thin continuous curves in 
Fig.~\ref{fg:deformed} a-c for different critical polar angles separating the 
two fission modes, for all three systems. As can be seen from the reasonable 
agreement of the mixed $\sigma_{m}^2$ values  with 
the observed fission fragment mass widths, we can phenomenologically explain 
the observed increase in the widths of the mass distributions when 
energy is decreased. It is interesting to note that the fusion-fission process 
is clearly dominated by the normal process at above Coulomb barriers and the 
"anomalous" fission process is  dominant at lower energies. However, 
experimental evidence suggests that the variations of mass distributions 
with excitation energies are similar  for the both processes, probably 
dominated by macroscopic forces, but differing quantitatively due to 
microscopic effects.

Extensive calculations of the multidimensional potential energy surface have 
successfully explained spontaneous and low energy fission phenomena 
\cite{NatureMoller01,PRLMoller04}. Calculated paths through the minimum 
energy valleys and over ridges in the potential surface showed that apart 
from the deformations and necking of the two nascent fragments, the 
left-right mass asymmetry also plays a crucial role. All the heavier than 
actinide nuclei show mass symmetric ($\alpha =0$) and mass asymmetric 
($\alpha \not= 0$) saddle shapes with a ridge separating the two down the 
scission path. The relative heights of the two saddles and the separating 
ridge governed the symmetric, asymmetric or a mixture of the two fission 
paths in specific cases. Recent extensions \cite{NPAIwamoto04,NPAAritomo04} of 
the five dimensional energy landscapes for fusion of $^{48}$Ca with  
$^{244}$Pu  have been carried out. In addition to the calculated  
minimum energy path  to reach the fusion meadow and the subsequent descent 
to the fission valley over a mass symmetric unconditional saddle corresponding 
to the fusion-fission (FF) path, at higher excitations, most of the paths may 
deviate through a mass symmetric saddle shape before fusion to re-separate in 
a quasi-fission (QF) reaction mode. In a very similar situation, in case of fusion of spherical 
projectiles with deformed $^{232}$Th nuclei, above the Coulomb barrier, the 
system follows a fusion-fission path over the mass symmetric unconditional 
saddle. But as the energy is decreased, these paths are progressively blocked 
and then the microscopic effects come into play. For the polar region of the 
deformed target, the system starts from an initial condition with varying 
deformation, separation and damping of radial motion. This results in the 
system finding a minimum energy path skirting the fusion meadow and over an 
almost mass symmetric saddle. In analogy to  skiers coming down a mountain 
slope from different heights (initial energy), go over a peak( fusion barrier) 
to a meadow (fusion) and continuing to slide over a small hillock (unconditional fission barrier) to reach the valley below (scission) in the established 
route (FF), those who start just below or at the peak, the normal route is 
blocked. However, if mountainsides are different (microscopic effect due to 
deformation) and a ridge exists near the peak, some of the skiers can 
reach the ridge and follow it  over a hilltop (conditional mass symmetric 
saddle) and reach almost the same spot at the valley in different route (QF).
 However, for a spherical target, the mountain sides are all similar and no 
ridges exist. The current experimental results strongly indicates the 
likely scenario described above and calls for detailed calculations of the
 energy diagrams for the motion of the nucleons through the dissipative system 
with different initial conditions.

We have clearly established with the present string of precise measurements  
that widths of the mass distributions  is a sensible tool to observe 
departure from the normal fusion-fission path in the fusion of heavy nuclei. 
The exact mechanisms for the departure  from normal fusion-fission paths are 
not known accurately, although it has been stressed that effect of any admixture of transfer fission can be ruled out. However, macroscopic effects such as the 
direction of mass flow or the mass relaxation time being too prolonged may 
not be the cause. It has been 
established earlier from the experimental barrier distributions, the reaction
 cross sections in $^{19}$F, $^{16}$O, $^{12}$C + $^{232}$Th in near and below 
Coulomb barrier energies are mostly for impact of the projectiles on the polar 
regions of the thorium nuclei. Following the quantum mechanical effects 
favouring  similar shapes in entrance and exit channels 
\cite{NatureMollerNV03}, we modify the simple postulation of the microscopic 
effects of the relative  orientation of the projectile to the nuclear symmetry 
axes of the deformed target \cite{HindePRL95}. We assume that for the 
non-compact entrance channel shape, the impact of the projectile in the polar
 region of $^{232}$Th target drives the system to an almost mass symmetric 
saddle shape, rather than a compact equilibrated fused system. The observed 
fragment mass widths can be quantitatively explained under such assumptions. 
At sub-coulomb barrier energies, the cross sections for quasi fission 
channel increases with increased initial separation of the two nuclei in 
touchy condition, i.e., the effect increased with the mass of the composite 
system. So quasi-fission is more prominent in $^{19}$F+ $^{232}$Th 
than $^{12}$C + $^{232}$Th. However in each system, as beam energy is 
increased, the reaction rapidly spread over all the nuclear surface and 
quasi-fission channels are overshadowed by normal fusion-fission. 
The above postulation is supported by the observation that for the 
spherical target $^{209}$Bi, where entrance channel compactness of shape is 
same for all relative target-projectile orientations, only normal 
fusion-fission paths, as characterized by the smooth variation of fragment 
mass widths with excitation energy, are observed. It is also worthwhile to 
note that effect of the anomalous mass widths increases with left-right mass 
symmetry in the entrance channel in case of $^{19}$F, $^{16}$O, $^{12}$C + 
$^{232}$Th system in consonance with our description. Our measurements indicate that higher entrance channel mass asymmetry and energies close to the Coulomb 
barrier are preferable to increase the probability of reaching the 
fusion meadow in synthesis of super-heavy elements in heavy ion reactions. 

Authors would like to thank the staff at NSC Pelletron for providing excellent 
beam and other logistical support and help during the experiment. Help of 
Drs. A. Saxena, D. C. Biswas, S.Chattopadhyay, Mr. P. K. Sahu during the 
experiments and  discussions with Drs R.K.Bhowmick and S.K.Datta 
are gratefully acknowledged. We are sincerely thankful to Prof. Binayak Dutta 
Roy for critical reading of the manuscript.




\bibliographystyle{unsrt}

\begin{thebibliography}{99}
\bibitem{NatureMollerNV03} P. Moller and A.J. Sierk, Nature {\bf 422} (2003)
        485 .
\bibitem{NishioPRL04} K. Nishio {\em et al.}, Phys.\ Rev.\ Lett.\ {\bf 93}(2004)
        162701.
\bibitem{HindePRL95} D.J. Hinde {\em et al.}, Phys.\ Rev.\ Lett.\ {\bf 74}(1995)
        1295.
\bibitem{myRAPID2} T.K. Ghosh {\em et al.}, Phys.\ Rev.\ C {\bf 69} (2004) 
        011604(R).
\bibitem{HalpernStrutinsky} I. Halpern and V. M. Strutinsky, in {\sl Proc. 
         of 2nd International Conference on Peaceful Uses of Atomic Energy, 
        (United Nations Publication, Geneva, 1958), Vol. 15, p 408}.
\bibitem{NMPRL96} N. Majumdar {\em et al.}, Phys.\ Rev.\ Lett.\ {\bf 77} (1996)
        5027.
\bibitem{RamPRL90} V.S. Ramamurthy {\em et al.}, Phys.\ Rev.\ Lett.\ {\bf 65}, 
        (1990) 25 .
\bibitem{ZhangPRC94} H. Zhang {\em et al.}, Phys.\ Rev.\ C {\bf 49} (1994) 926.
\bibitem{SamantEPJ00} A.M. Samant {\em et al.}, Eur.\ Phys.\ J.\ A.{\bf 7} 
        (2000) 59 .
\bibitem{KailasPhysRep} S. Kailas , Phys.\ Rep.\ {\bf 284} (1997) 381 .
\bibitem{Abe} M. Abe, KEK Report No. 86-26, KEK TH-28, 1986.
\bibitem{PRC87Shen} W.Q. Shen {\em et al.}, Phys.\ Rev.\ C {\bf 36} (1987) 115 .
\bibitem{myNIM04} T.K. Ghosh {\em et al.}, Nucl.\ Instr.\ and Meth.\ {\bf A 540}         (2005) 285.
\bibitem{PB95} P. Bhattacharya {\em et al.}, Nuovo\ Cimento\ Soc.\ Ital.\ Fis.,\         A {\bf 108}, (1995) 819, D.J. Hinde {\sl et al.}, Phys.\ Rev.\ C 
        {\bf 53} (1996) 1290
\bibitem{myRAPID1} T.K. Ghosh {\em et al.}, Phys.\ Rev.\ C {\bf 69} (2004)
         031603(R).
\bibitem{LestonePRC97} J.P. Lestone {\em et al.}, Phys.\ Rev.\ C {\bf 55} (1997)         R16 .
\bibitem{NMPRC95} N. Majumdar {\em et al.}, Phys.\ Rev.\ C {\bf 51} (1995) 
         3109 .
\bibitem{BackJH} B.B. Back {\em et al.}, Fission at Sub-barrier energies, 
         presented at 6th Winter Workshop on Nuclear Dynamics,(Jackson Hole, 
         Wyoming,USA, 1990) 
\bibitem{NatureMoller01} P. Moller, D.G. Madland, A.J. Sierk and A. Iwamoto, 
        Nature {\bf 409} (2001) 785 .
\bibitem{PRLMoller04} Peter Moller, Arnold J. Sierk and Akira Iwamoto, 
        Phys.\ Rev.\ Lett.\ {\bf 92} (2004) 072501 .

\bibitem{NPAIwamoto04} A. Iwamoto {\em et al.}, Nucl.\ Phys.\ A {\bf 738} 
        (2004) 499.  
\bibitem{NPAAritomo04} Y. Aritomo {\em et al.}, Nucl.\ Phys.\ A {\bf 744} 
        (2004) 3. 
\bibitem{NMPrivate} N.Majumdar (private communication)
\bibitem{Vul} E. Vulgaris {\em et al.}, Phys.\ Rev.\ C {\bf 33}(1986) 2017 .

\end{thebibliography}

\end{document}